# Silicon Photonics DWDM NLFT Soliton Transmitter Implementation and Link Budget Assessment

A. Moscoso-Mártir[1*], J. Koch[2], J. Müller[1], S. Sharif Azadeh[1,3], S. Pachnicke[2], J. Witzens[1]
[1] Institute of Integrated Photonics, RWTH Aachen University, Alois-Riedler-Str. 4, 52074 Aachen, Germany
[2] Chair of Communications, Faculty of Engineering, Kiel University, Kaiserstraße 2, 24143 Kiel, Germany
[3] Now at NINT Department, Max Planck Institute of Microstructure Physics, 06120 Halle, Germany
e-mail: amartir@iph.rwth-aachen.de

**ABSTRACT**

We analyze the implementation and capabilities of a silicon photonics transmitter capable of modulating and multiplexing groups of four solitons with different frequency and time spacing based on the nonlinear Fourier transform.

**Keywords**: Nonlinear optics, DWDM system, NLFT system, integrated optoelectronic circuits.

## 1. INTRODUCTION

The optical backbone infrastructure has to support the continuing demand for exponentially growing data throughput. A cost-effective solution to increase the channel capacity avoiding the deployment of new fibers consists in increasing the limits of useable power per fiber without incurring nonlinear penalties [1]. In transmission systems based on the nonlinear Fourier transform (NLFT), the nonlinear Kerr effect simply results in a change of the phase in the nonlinear spectrum that can be compensated at the receiver (Rx) end [2]. However, current systems present some limitations in their ability to create higher-order solitons for more efficient data transmission [3]. These limitations stem from the required computational complexity, high sampling rate and high bandwidth (BW). To overcome these, it has been proposed to shift the soliton merging from the electrical to the optical domain using a photonic integrated circuit (PIC) [4]. In this solution, first-order solitons are generated on individual optical carriers and are multiplexed later using a programmable architecture that allows a precise control over time and frequency spacing between the individual solitons with minimal spectral distortion [4]. Since only first-order solitons have to be generated and detected in the electrical domain, BW requirements on the electrical and optoelectronic components are significantly reduced. Moreover, it is possible to parallelize first-order soliton generation and detection, further reducing the digital signal processing (DSP) speed requirements.

In this paper, we present the realization of a silicon photonics (SiP) transmitter (Tx) that can modulate four first-order solitons and multiplex them with adjustable time-delays by means of a programmable delay network implemented with coupled (ring-)resonator optical waveguide (CROW) optical add-drop multiplexers (OADMs) [5]. Based on the modeled and measured characteristics of the fabricated PIC, and previous experience with fully functional wavelength division multiplexed (WDM) transceivers [6], a complete model for a system setup combining commercially available equipment and the proposed PIC is established and analyzed. A link budget is derived to obtain realistic power values for the multiplexed soliton pulses at the output of the PIC. This information is used to upgrade the simulation tool presented in [4] to model the link together with the fabricated chip, so that the features and programmability of the PIC can be investigated. Using this simulation tool, we have run different scenarios concerning pulse timing and explored the capabilities of the SiP DWDM NLFT soliton transmitter under development. Experimental system-level results available by the time of the conference will also be presented.

## 2. TRANSMITTER IMPLEMENTATION AND LINK BUDGET

The envisioned system as well as a micrograph of the recently delivered system chip fabricated in a multi-project wafer run at Advanced Micro Foundries (AMF) are shown in Fig. 1 and Fig. 2(a). A comb source from Pilot Photonics (Lyra-OCS-1000) is used to provide the carriers for the first-order solitons. The comb has an adjustable free spectral range (FSR) of $10 \pm 4$ GHz that can be used to adjust the soliton frequency spacing. Each line has a power of -6 dBm and a linewidth of 80 kHz. Four of these lines are filtered with an external 45 GHz passband optical filter with 2 dB insertion loss (IL). The selected carriers are amplified to 10 dBm with an EDFA with a noise figure (NF) of 5 dB, so that the total power injected in the PIC does not exceed 16 dBm to prevent permanent damage to the grating coupler (GC, IL = 3 dB) used as a coupling device. Inside the chip, four $2^{nd}$-order CROW OADMs (IL = 1.6 dB, BW = 6.5 GHz) route carriers to one of the two input ports of two IQ Mach-Zehnder modulators (MZM), as shown in Fig. 2(a). The designed IQ-MZMs with slotted transmission lines and inter-phase-shifter cross-talk suppression [7,8] have 4.4 mm long phase shifters, a BW of 14 GHz optimized for the present application, a $V_\pi L$ of 2.45 V·cm and 4.5 dB IL. Operated with 1 $V_{pp}$ signals and biased to achieve full extinction during soliton shaping, they introduce an insertion loss and modulation penalty of 13.5 dB, defined as the attenuation of the peak power after modulation. Afterwards, channels 1 and 2, that respectively carry the same information as channels 3 and 4, are delayed (IL = 3 dB) as a means to emulate four independent channels. All channels are multiplexed onto one out of two bus waveguides using four 4th-order CROW OADMs (IL = 2 dB, BW = 17.5 GHz) that are tuned to achieve equal peak power for all channels.

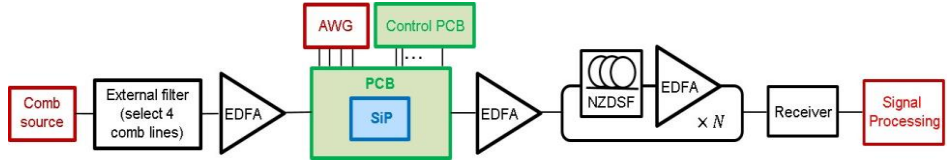

*Figure 1. Block diagram of investigated link.*

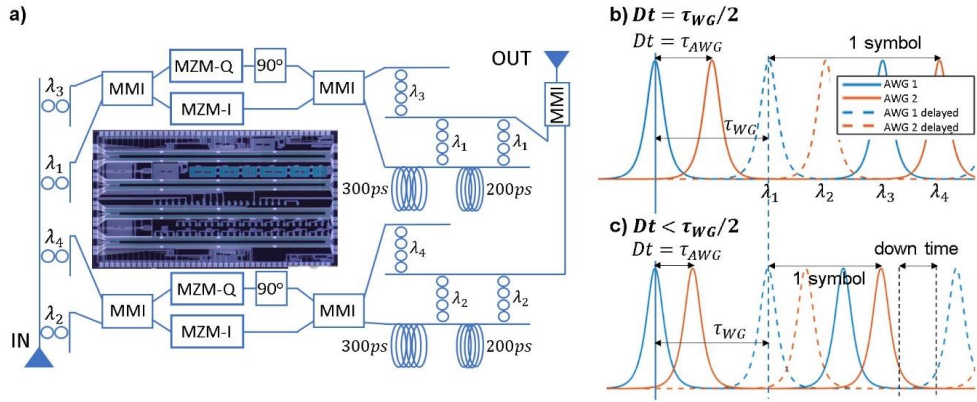

*Figure 2. (a) PIC block diagram and chip micrograph (inset). (b,c) Soliton pulse sequencing within the transmission window.*

A multimode interferometer coupler (MMI) finally serves to combine the two bus waveguides into one (IL = 3 dB), that is then routed off chip via a second GC. Merging via an MMI allows for pulses of even and odd channels to have partial spectral overlap without clipping of the spectra [4]. Including interconnection and monitor tap losses (1.5 dB), we obtain a soliton peak power of -20.6 dBm per channel. These power levels have to be reamplified with an EDFA to reach launch powers that fulfill the soliton condition, depending on soliton pulse width and the length of the used recirculating, amplified non-zero dispersion shifted fiber (NZDSF) loops [1].

In the implemented transmitter [Fig. 2(a)], neighboring channels (single increment in channel number also corresponds to closest spectral and time overlap inside the transmission window (TW), since $\lambda_1 < \lambda_2 < \lambda_3 < \lambda_4$ and $t_1 < t_2 < t_3 < t_4$) are sent to different IQ-MZMs to ensure fully independent data streams. This also ensures that channels multiplexed onto the same bus waveguide at the output of an IQ modulator also have sufficiently low spectral overlap to be routable by the OADMs with minimal cross-talk. The four optical channels are modulated with only a 4-channel arbitrary waveform generator (AWG). The delay $\tau_{WG}$ applied to channels 1 and 2 by on-chip delay loops can be selected to be either 300 ps or 300 ps + 200 ps. Together with the pulse repetition time at the output of the AWG (duration of a complete symbol, i.e., a 4-pulse TW) as well as the time delay between independent pulse trains applied to the two IQ-MZMs ($\tau_{AWG}$), this allows reprogramming of the pulse-to-pulse time delays (Dt). For example, this allows to explore pulse spacings Dt of 250 ps or 150 ps with pulse trains fully filling 1000 ps or 600 ps long TWs [Fig. 2(b)]. A 3rd configuration also modeled below consists in closer spaced pulses (Dt = 100 ps) only partially filling a 500 ps long TW also including a 100 ps down time [Fig. 2(c)].

## 3. SIMULATION SETUP AND RESULTS

For the simulations, blocks of 40,000 bits are QPSK modulated and split up into two channels. Using these symbols, soliton pulses are created with a characteristic pulse width ($T_0$) of 38 ps, a BW of 8.8 GHz (99% power) and a peak amplitude of 1 $V_{pp}$. These soliton pulses are then applied to the IQ-MZMs to modulate and multiplex the four channels. The relative carrier frequencies are set to $\Delta f$ = [-15 GHz, -5 GHz, 5 GHz, 15 GHz]. As mentioned above, soliton pulses have a peak power of -20.6 dBm at the output of the PIC, which needs to be amplified by an EDFA (NF = 5 dB) up to -0.3 dBm fulfilling the soliton condition. After amplification, the combined solitons are launched into N spans of 50 km NZDSF combined with an EDFA (NF = 5 dB), resulting in a 28 dB optical signal-to-noise ratio (OSNR) at the Rx after 3,750 km (OSNR accounting for 4 soliton pulses and noise in the entire TW). On the Rx side, the 4-pulse trains are wavelength demultiplexed and sent to four parallel coherent Rx. After demodulation, each soliton pulse is low-pass filtered, sampled and digitized. Finally, solitons are normalized and analyzed using the forward-backward method, followed by a blind phase search for phase recovery [4].

We have simulated the system to understand how the time spacing affects soliton integrity and demodulation. As seen in Fig. 3(a), it is possible to transmit groups of 4 solitons with Dt = 250 ps (8 Gbps) up to 2,000 km with a BER below the hard-decision forward error correction (HD FEC) limit and up to 2,750 km below the soft-decision (SD) FEC limit. Similar results are obtained for a tighter soliton sequencing with Dt =150 ps (13.3 Gbps). Packing the solitons even closer clearly affects their integrity as can be seen for Dt = 100 ps. This shows that the four solitons can be packed together in a TW of width $4T_0$ with small penalty, but that higher compactification is detrimental. Ripples are seen in the BER curves at 1950 km for Dt = 150 ps and at 3500 km for Dt = 250 ps. In soliton merging, soliton pulses propagate together with different group velocities, changing their initial position within their TW. At specific distances they can partially or even completely overlap, as seen in Fig. 3(b). This makes it more difficult to recover them at the Rx even when compared with longer distances where they are again equally spaced [Fig. 3(c)], as a consequence of the partial spectral overlap of the soliton pulses [4] making wavelength demultiplexing imperfect. Changing the frequency spacing or the initial time domain separation as enabled by the Tx reconfigurability allows shifting these ripples away from a targeted transmission length. Hence, this problem can be avoided in a real transmission scenario. While data rates that are currently considered are still low, they will be increased once the system has been experimentally proven by increasing the number of solitons, using more complex modulation formats and reducing the width $T_0$ of the solitons. With the current chip, we could generate QPSK modulated solitons up to 20 Gbps. It is also possible to redesign the IQ-MZM to achieve higher BWs [8] and higher values of Vpp could be used to reduce the modulation penalty. Furthermore, data dependent time shifts on a small scale can be included to introduce amplitude modulation in the NLFT domain [1].

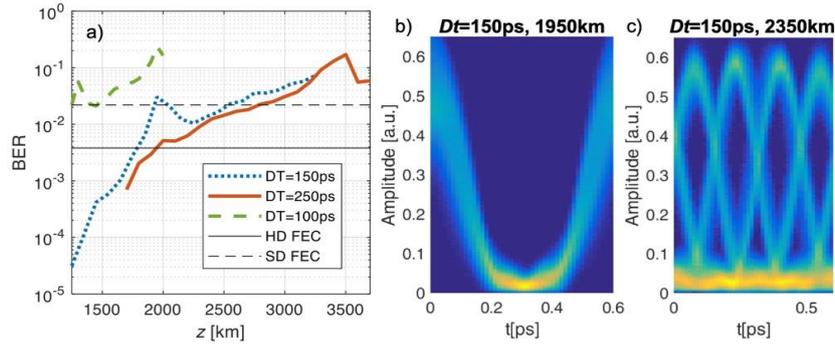

*Figure 3. (a) BER as a function of distance for different Dt. Soliton pulse eye diagrams for (b) Dt = 150 ps at 1,950 km and (c) Dt = 150 ps at 2,350 km.*

## 4. CONCLUSIONS

We have presented a SiP DWDM soliton Tx that has been characterized and is currently being used in experimental system tests. It can modulate four first-order solitons and multiplex them with different frequency and time spacing. Nonlinear link simulations have been run for different pulse-to-pulse spacing and fiber lengths. Simulation results show that solitons can be compactified together in a transmission window $4T_0$ and still allow to reach up to 2,500 km with a BER below the SD FEC limit. Furthermore, we have explored the evolution of soliton merging over distance, providing a solution to soliton overlap by shifting it out of the region of interest via Tx reconfiguration. There are multiple paths towards increasing the data rates we are presently working with that will be taken once the current configuration has been experimentally proven.